# Medical articles in questionable journals are less impactful than those in non-questionable journals but still extensively cited

Dimity Stephen

German Centre for Higher Education Research and Science Studies (DZHW), Schützenstrasse 6A, Berlin, Germany

stephen@dzhw.eu, ORCID: 0000-0002-7787-6081

A key feature of questionable journals is a lack of adequate peer review of their articles. Content of thus unknown quality may be utilised by unsuspecting practitioners or incorporated into peer-reviewed research, becoming legitimised. It is therefore necessary to examine the citation patterns of articles in questionable journals to understand the impact and reach of research in questionable journals. Similar research has tended to focus on authors from low- and middle-income countries. As such, this study investigates the profile and impact of research in questionable journals by authors in Germany. Questionable journals were identified by matching journals with articles by authors at German institutions from Dimensions to Cabell's Predatory Reports. Metadata for these articles and a comparative sample of articles in non-questionable journals were extracted from Dimensions and the 3 year citations, self-citations, uncited rate, profile of co-authoring and citing countries, and institution type of authors were compared between groups. Nearly 600 articles in 88 questionable journals were published by German authors in 2010-2020. Three-quarters were in the medical and health sciences. Medical articles in questionable journals received significantly fewer citations than similar articles in non-questionable journals. However, articles in questionable journals were still extensively cited in 1,736 primarily non-questionable journals. Self-citations accounted for only 12% of these citations. Authors from non-university medical facilities were over-represented in articles in questionable journals. System-level changes are necessary to eliminate questionable journals and shift high-quality research into reputable venues.

**Key words:** questionable journals; questionable publishing; predatory publishing; research integrity; citation analysis

**Statements and Declarations**

**Funding information**:

No funding was received for conducting this study.

**Competing interests:**

The author has no relevant financial or non-financial interests to disclose.

**Introduction**

Developments in publishing practices and academic culture have brought with them new challenges. The "publish or perish" culture exerts pressure on academics, particularly those early in their careers, to regularly publish their research in order to be competitive in obtaining funding and perform well in evaluations (Kurt, 2018; Richtig et al., 2018). Consequently there is high demand for publishing outlets, particularly those offering fast publishing times. Simultaneously, technological advancements and cultural shifts towards greater inclusivity in research have introduced the Open Access (OA) publishing model (Richtig et al., 2018). In OA publishing, instead of publishers being financed by subscriptions to their journals, authors or their institutions pay the publishers Article Processing Charges (APCs) to publish their papers as freely available online content (Björk et al., 2020; Richtig et al., 2018). OA publishing is fundamentally a positive movement toward inclusivity in academia, facilitating legal access to research for anyone with an internet connection. However, the combined effects of the publish or perish culture and OA publishing has allowed questionable journals to flourish in recent years (Richtig et al., 2018).

Questionable journals, also known as predatory journals, are journals that demonstrate "dishonest tendencies, lack scientific rigor, and function primarily for financial gain" (Kurt, 2018). Under the guise of OA, publishers of questionable journals may charge authors APCs, and sometimes additional or exorbitant fees (Kurt, 2018), to publish in their journals. The short times to publication typically offered by these journals appeal to researchers under pressure to produce (Kurt, 2018). However, these journals do not afford researchers the benefits of publishing in a reputable journal, such as effective peer review and visibility to a relevant audience. Instead, questionable publishers have been observed to spam researchers for submissions and editorial positions, falsely claim impact metrics and/or indexation in respected databases, and list individuals as editorial staff without their knowledge or consent (Björk et al., 2020; Bohannon, 2013; Kurt, 2018). However, perhaps most disconcertingly, questionable publishers typically do not conduct adequate peer review. This has been highlighted by so-called "sting operations" in which bogus articles are submitted to potentially questionable journals to see if they are accepted. The quality of the articles is intentionally so poor that acceptance can only mean that ineffective peer review was conducted. For instance, Bohannon (2013) submitted 304 versions of a falsified medical science manuscript to 255 OA journals, of which 157 accepted the paper, despite its "experiments [being] so hopelessly flawed that the results are meaningless". Sixty percent of the journals made a publishing decision without any evidence of peer review being undertaken, and 16 manuscripts were accepted in contradiction to the direct advice of peer reviewers (Bohannon, 2013). Evidently, ineffective peer review is a notable concern with questionable journals.

Publishing in questionable journals is thus damaging for both the authors – content is typically not indexed in reputable databases thereby potentially reducing its impact and squandering funding spent on APCs – and the broader community as unreviewed research becomes publically available with credibility lent by its publication in an apparently academic journal. Spurred by events such as the COVID-19 pandemic and the anti-vaccine movement, there has recently been more interest from the general public in scholarly articles. While such consumption should be encouraged, individuals without the necessary experience to assess a publishing venue or an article's content can access and perpetuate potentially damaging information published in questionable journals. For instance, in one study 74% of oncologists reported being challenged by their patients with scientific literature, 14% of which was from questionable journals (Richtig et al., 2019). However, even training may not be sufficient to prevent the uptake of such content: the same study found that 30% of surveyed oncologists in Germany and Austria were unaware of the concept of predatory journals, and around half felt they could not identify one (Richtig et al., 2019). This is particularly concerning as over 95% of the practitioners reported using journal articles to inform their clinical treatment decisions (Richtig et al., 2019) and 5% of nursing studies in questionable journals contained findings that could be potentially harmful to patients (Oermann et al., 2018). More indirectly, articles that have not undergone adequate peer review can be cited by articles in non-questionable journals and become legitimised. Consequently, research that, for example, does not adhere to medical research reporting guidelines, adequately describe methods, consider risks of bias, review relevant literature, or receive approval from an ethics committee for human or animal research (Moher et al., 2017; Oermann et al., 2018) can enter mainstream academia. It is therefore necessary to examine the citation patterns of articles in questionable journals and the characteristics of those citations to understand the impact and reach of research questionable journals.

Previous studies have found substantial variation in the average citations of articles in questionable journals. For example, Bagues et al. (2019) examined citations for the 5,798 articles from Beall's list journals they identified in Italian researchers' CVs. The articles' median citations were low at 3 and 23% had never been cited, which increased to 33% when self-citations were excluded. However, the 10% most cited articles received 20 or more citations each and one article was cited 399 times, indicating some articles in questionable journals can draw substantial attention. Similar average citation levels were observed by Ezinwa Nwagwu & Ojemeni (2015), who found an average of 2.3 citations per paper for 5,601 articles in 32 questionable biomedical journals from two Nigerian publishers, while Björk et al. (2020) observed an average of 2.6 citations within 5 years of publication of 250 articles in questionable journals. Moussa (2020) observed higher rates of a mean of 6.2 citations per paper for a sample from ten questionable marketing journals that had mimicked the names of reputable journals, and the most cited articles received 40-217 citations. However, a sample of 3,427 articles published in the questionable journals from Turkey received on average fewer than 0.5 citations per article (Akça & Akbulut, 2021) and in a random sample of 124 journals from Beall's list, Frandsen (2017) observed 1,295 citations within 3 years of publication, giving an average citation rate of just 0.04 citations per article.

A small number of studies have compared the citations of articles in questionable journals against relevant control groups. These studies have tended to find that articles in questionable journals typically remained uncited more often and received substantially fewer citations than similar articles in non-questionable journals (Akça & Akbulut, 2021; Björk et al., 2020; Moussa, 2020). For instance, 250 articles in questionable journals received an average 2.6 citations and 56% remained uncited compared to 18.1 citations on average and 9% uncited rate for a sample of 1,000 articles from Scopus-indexed journals (Björk et al., 2020). The 10,935 citations garnered by 1,246 articles in questionable marketing journals constituted only 3-7% of the citations obtained by articles in the equivalent reputable journal, while the 0.5 citations per article observed for questionable journals from Turkey was lower than non-questionable journals in the same fields (Akça & Akbulut, 2021). In another study, 58 questionable accounting journals received on average 3.8 citations per article, placing them equivalent to the 11th percentile of 61 Scopus-indexed journals. However, the eight most cited journals were cited similarly to the control group of Scopus journals (Walters, 2022). The citation-based impact of articles in questionable journals varies depending on the characteristics of the sample, such as field and country, but overall, citation rates tend to be relatively low and typically less than similar articles in non-questionable journals. However, some articles and journals achieve high citations and are cited similarly to non-questionable journals.

In terms of reach, studies have often found researchers in Africa, South Asia and South-East Asia to constitute both the majority of authors and citing authors of articles in questionable journals (Ezinwa Nwagwu & Ojemeni, 2015; Frandsen, 2017; Kulczycki et al., 2021). Frandsen (2017) determined that citing authors were often inexperienced, with around half having no previous publications or citations in Scopus themselves. However, citing authors from Europe and North America were more experienced researchers in terms of prior publications and citations than citers from Africa, South and South-East Asia. Also, researchers in the USA, Sweden, and Australia most frequently both authored articles in questionable nursing journals and also cited these articles from non-questionable journals in Scopus (Oermann et al., 2019), demonstrating both the global reach of articles in questionable journals and their penetration of mainstream, peer-reviewed research channels. Another study found that nearly 90% of 68 reputable marketing journals had been "contaminated" by citations to articles in four questionable journals (Moussa, 2021). R. Anderson (2019) also examined the citations of seven questionable medical journals and found there was a relatively low rate of citations at just 1-40 citations in Web of Science (WoS) or ScienceDirect articles to five of the journals. However, this represented up to 37% of the questionable journals' articles being reflected in mainstream literature. Similarly, Ross-White et al. (2019), Kulczycki et al. (2021), and Oermann et al. (2019) found that, although only portions of their respective samples of biomedical, social science, and nursing articles in questionable journals had been cited, these citations represented penetration into 157 systematic reviews, 2,338 WoS-indexed journals, and 141

Scopus-indexed journals. Evidently then, although citations of articles in questionable journals tends to be low, not inconsequential proportions of research of unknown quality in questionable journals is being perpetuated via a vast number of reputable publishing outlets.

The aim of this study was to investigate the profile and citation impact of research in questionable journals authored by researchers at German institutions. Research on questionable journals often focuses on low- and middle-income countries as authors from these countries often constitute the majority of authors in questionable journals (e.g. Xia et al. [2015]; Ezinwa Nwagwu & Ojemeni [2015]; Frandsen [2017]). However, more than 5,000 researchers in Germany recently published in questionable journals (Offord, 2018) and up to 50% of authors in questionable journals are from high- or upper-middle-income countries (Cohen et al., 2019; Moher et al., 2017), indicating authors in these countries may also engage in or be susceptible to questionable publishing practices. Further, high-income European countries have a citation advantage over other countries, even when publishing in the same journals (Akre et al., 2009). Consequently, German research may be more readily cited than research from low- or middle-income countries, despite its publishing venue, extending the reach of works in questionable journals. As such, it is important to understand the profile and citation impact of German research in questionable journals, an examination which to date appears not to have been undertaken. Therefore, this study compared the citations and countries of co-authors and authors citing German-authored articles in questionable journals against a comparable sample of German-authored articles in non-questionable journals to examine i) the extent to which German-authored articles in questionable journals are cited, ii) the profiles of authors and journals citing this research, and iii) whether the citation impact and profile varies from research in non-questionable journals.

**Methods**

Questionable journals were defined as those indexed in Cabell's Predatory Reports (CPR). Launched in 2017, CPR is a subscription-based database of over 16,000 journals classified via manual review as "predatory" based on more than 60 indicators of publishing practices, such as peer review and metrics usage.[1] I selected CPR as the basis for questionable journals it is current and transparent in its justification for classifying journals as questionable.

To identify German research in questionable journals, I first identified all journals in which authors at German institutions published between 2010 and 2020 in the Dimensions bibliometric database (20,156 journals). Dimensions was used as the basis for bibliometric data because it contains a broader sample of research than the more restricted, quality-controlled content in Scopus and WoS (Visser et al., 2021). Data from Dimensions were supplied by Digital Science as a snapshot of data up to April 2021. As the comparison between journals in Dimensions and CPR was to be largely manual, I first reduced the sample of journals by excluding those listed in the Directory of Open Access Journals (DOAJ) and in WoS as these journals are unlikely to be questionable. The DOAJ is an index of more than 17,500 peer-reviewed, OA journals across all fields, languages, and countries, for which journal publishers apply and are reviewed for inclusion.[2] Metadata for journals in the DOAJ were obtained from the DOAJ's public data file[3] for January 25, 2022. Data from WoS were obtained from the German Competence Network for Bibliometrics'[4] in-house version of the database, which covered data until April 2021.

After removing punctuation and converting all titles to lowercase, I matched journal titles from Dimensions to those in WoS and the DOAJ using the fuzzyjoin package (Robinson, 2020) in R (R Core Team, 2020) and a threshold Levenshtein distance of 6 (Levenshtein, 1966). Matches with distances between 3 and 6 were manually confirmed, and I removed from the sample 4,066 journals in DOAJ and 10,214 journals in WoS. Of the remaining 7,389 journal titles, I excluded a further 5,134 that were published by reputable publishing houses such as Wiley, Springer, or De Gruyter. I searched the remaining titles in CPR on 8 and 9 February 2022 to identify questionable journals and for all matches I recorded the number and type of violations provided for classifying the journal as questionable.

---

[1] https://www2.cabells.com/predatory

[2] https://doaj.org/about/

[3] https://doaj.org/docs/public-data-dump/

[4] https://bibliometrie.info/en/

I then extracted from Dimensions the metadata of all articles with German authors published 2010-2020 in journals that appeared in CPR, including all authors' names, institutional affiliations, the article's discipline, and its 3-year and total citation counts. I used a 3-year citation window as articles in both questionable and reputable journals tend to accrue a level of citations representative of their long-term impact within 1-4 years of publication (Akça & Akbulut, 2021; Oermann et al., 2019; Wang, 2013). I also extracted from Dimensions the metadata of all articles citing these articles, including authors' names, country-level affiliations, and the publishing journal. I also searched the journals of citing articles in CPR to determine if they too were questionable. To identify instances of self-citation, I cleaned the authors' and citing authors' names of punctuation and merged the first and surnames and then fuzzy matched these names using a threshold Levenshtein distance of 8. All matches were manually checked for accuracy. Self-citations were calculated on the basis of any author of the article in the questionable journal citing themselves, not just the author affiliated with the German institution. To analyse the institution types of authors of articles in questionable journals, I attributed one of the following types to each author in Germany based on their affiliation address string: university, university hospital, non-university research institution, government agency, hospital or medical clinic, company, not-for-profit organisation, independent researcher, or missing information.

As a control group against which to compare the characteristics of articles in questionable journals, I randomly sampled from Dimensions 348 German-authored articles published in 2010-2018 in non-questionable OA journals that were classified as medical and health sciences (MHS) articles. Articles were restricted to MHS to ensure the effect of disciplinary citation practices was controlled for and MHS was the only discipline with an adequate number of articles for analysis in the questionable journals sample (348 articles published until 2018, allowing a 3-year citation window). OA journals were selected to ensure equivalence of a potential citation (dis)advantage of OA journals, as all questionable journals were OA. Eligible OA journals were identified via the previous step of matching journals with German-authored articles from Dimensions with the DOAJ list, as verified legitimate OA journals. Nature and PLoS journals were excluded as these highly cited, high-output journals are not generally representative of OA journals and could skew the citations of the control group. I then drew a random sample of 348 articles from journals classified as MHS. The same process of extracting metadata from Dimensions, identifying self-citations, and classifying institution types was also undertaken for this control group.

Finally, I compared the sample of articles in questionable journals and the control group on the profiles of co-authors' and citing authors' country affiliations, the percentage of citations that were self-citations, the percentage of articles that were uncited after 3 years, and the distribution of 3-year citations between groups. As expected, the distributions of citations in both samples were skewed toward 0 and therefore not normally distributed. As such, I used the non-parametric Mann-Whitney-Wilcoxon test to compare the groups' 3-year citations for significant differences.

**Results**

In total, there were 88 questionable journals with German-authored articles indexed in Dimensions. The mean number of violations CPR attributed to the questionable journals was 4.8 (range = 1-8, IQR = 3.8-6.0). The most common violations were in relation to publication practices (70, 79.5%), peer review (59, 67.0%), website issues (57, 64.8%), and access and copyright conditions (43, 48.9%). Between 2010 and 2020, 591 articles were published in these 88 journals by authors affiliated with German institutions. The peak occurred in 2014 and 2015 when 85 and 87 articles appeared. The number of articles declined sharply after 2017 to 19 in 2020, the lowest level over the time-series. This decline coincidences with a peak in informational articles about predatory publishing (Krawczyk & Kulczycki, 2021) and thus may reflect changes in publishing practices due to increased awareness of questionable journals. Articles in questionable journals represented less than 0.1% of all research produced in Germany annually. Three-quarters (449, 75.1%) of the articles were in MHS, followed by 13.2% (79) in natural sciences, 5.4% (32) in engineering and technology, 3.9% (23) in social sciences, 1.8% (11) in agricultural sciences, and 0.7% (4) in the humanities. A small number of articles were attributed to more than one field.

Analyses of citation counts and co-author and citing authors were based on the two samples of 348 MHS articles published between 2010 and 2018 in questionable and non-questionable journals. The articles in questionable journals received 3,481 citations from 3,407 distinct articles in 1,736 journals, while the articles in non-questionable journals received 8,846 citations from 8,825 distinct articles in 2,869 journals. Self-citations accounted for 12.3% (428) of the citations received by the articles in questionable journals, which was slightly lower than the rate of 15.6% (1,376) self-citations observed for articles in non-questionable journals. Nearly all citations to articles in questionable journals (3,293, 94.6%) were from non-questionable journals. Forty-two articles in questionable journals (12.1%) were uncited three years after publication, compared to 8.3% (29) of articles in non-questionable journals. The mean 3-year citations of articles in questionable and non-questionable journals were 4.6 and 10.2, respectively, while mean total citations were 10.0 and 25.4, respectively. The distributions of the 3-year citations of the articles in both groups are shown in Figure 1. Articles in questionable journals received significantly fewer citations within 3 years of publication than articles in non-questionable journals (W = 42,882, $p < 0.001$).

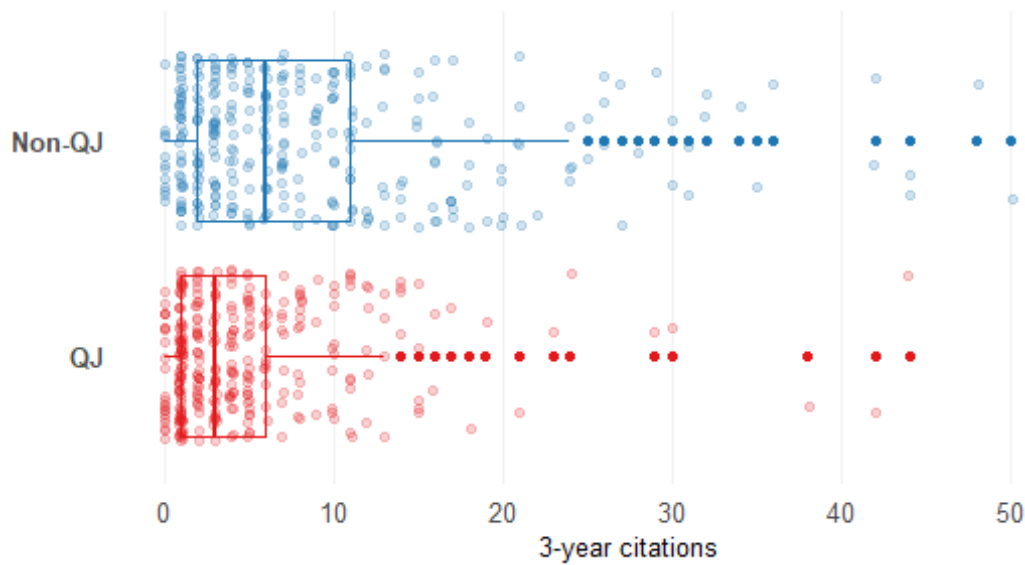

*Figure 1: Distributions of 3-year citations of MHS articles in questionable and non-questionable journals. Figure excludes 5 non-questionable articles with 64-247 citations.*

The left panel of Figure 2 shows the percentage of German-authored MHS articles in questionable and non-questionable journals that were co-authored with selected countries. These countries are those that co-authored more than 1% of articles in at least one of the journal groups. Israel, Ireland, and Saudi Arabia did not co-publish any articles in questionable journals and so therefore do not have data points for this group. Whole counting of co-authors was applied at the level of countries, therefore the percentages do not sum to 100%.

Notably, the countries with which German authors collaborated were similar between groups, although the extent of collaboration was typically reduced for articles published in questionable journals. For example, 17.5% of articles in non-questionable journals were co-authored with the United States, however an American co-author was only present on 7.8% of articles in questionable journals. Similarly large decreases were evident for the other key collaboration partners Switzerland, the United Kingdom, and the Netherlands. Co-authors from Belgium, Greece, and India were exceptions to this trend as they co-published slightly more articles in questionable than non-questionable journals.

Articles in questionable journals tended to be national outputs: only 24.7% (86) of these articles involved international collaboration compared to 57.5% (200) of articles in non-questionable journals.

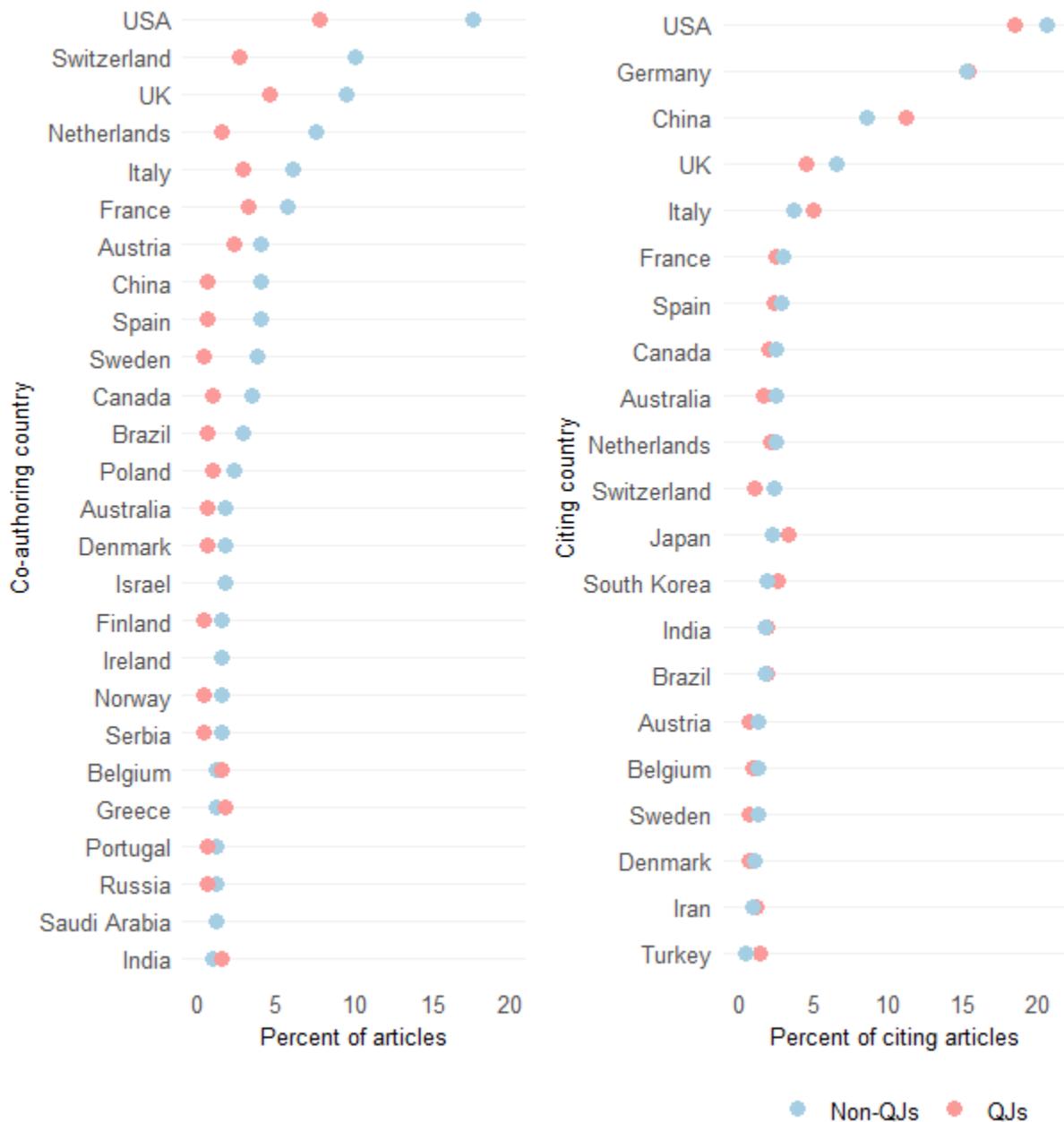

*Figure 2: Share of MHS articles in questionable and non-questionable journals co-authored by selected countries (left) and countries' fractional shares of citations to MHS articles in questionable and non-questionable journals (right).*

The right panel of Figure 2 shows selected countries' fractional share of articles citing German-authored MHS articles in questionable and non-questionable journals. These countries are those that accounted for more than 1% of citing articles in one or both of the journal groups. Fractional counting was applied at the level of the author and aggregated to country level. This panel excludes 5% and 7% of articles citing non-questionable and questionable journals, respectively, that did not have any information about the authors' affiliations. As with co-authorship, the pattern of countries citing German-authored articles was similar between groups. Only the United States, the United Kingdom and Switzerland displayed notable reductions in their share of citations between groups. For instance, the United States and the United Kingdom accounted for 20.6% and 6.4% of citations to non-questionable journals, which reduced to 18.5% and 4.4% of citations to questionable journals, while Switzerland's share dropped from 2.5% to 1.0%. Conversely, the share of citations to questionable journals was higher than non-questionable journals for China (11.2% and 8.5%), Italy (5.0%, 3.7%), Japan (3.2%, 2.2%), South Korea (2.5%, 1.8%), and Turkey (1.3%, 0.4%). Germany itself accounted for 15% of citations in each group.

Finally, the fractional shares of MHS articles in questionable and non-questionable journals by the authors' institution types are shown in Figure 3. Evident here again is the reduced international collaboration on articles in questionable journals with international co-authors accounting for 33.8% of articles in non-questionable journals but only 13.0% of articles in questionable journals. Universities and university hospitals accounted for the largest shares of articles in both journal types, which reflects the structure of the German academic system with universities and their affiliated facilities contributing the largest share of publications (Stephen & Stahlschmidt, 2021). Notably, however, authors from university hospitals and hospitals or medical clinics accounted for a substantially larger share of articles in questionable journals (38.9% and 13.0%) than non-questionable journals (16.6% and 1.4%). Conversely, authors from universities and non-university research associations accounted for less content in questionable journals (25.1% and 1.6%) than non-questionable journals (32.3% and 7.2%).

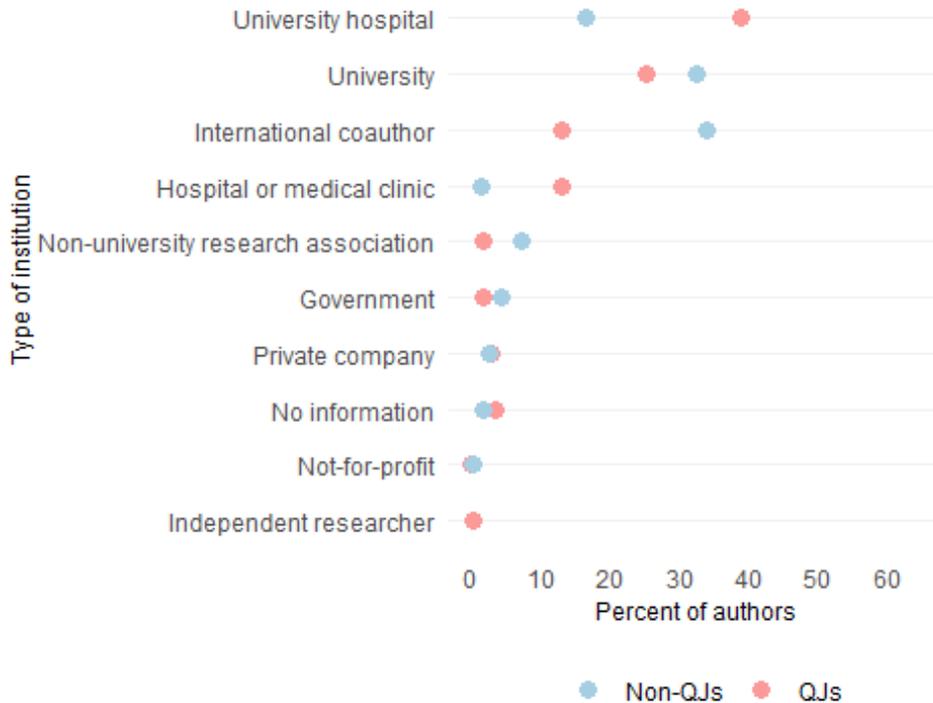

*Figure 3: Fractional shares of articles in (non-)questionable journals by authors' institution types.*

**Discussion**

This study identified 591 articles by researchers from German institutions published in 88 questionable journals between 2010 and 2020. Three-quarters of German research in questionable journals was in the medical and health sciences. However, this may reflect that a substantial proportion of research from both Germany (Stahlschmidt & Stephen, 2020) and indexed in Dimensions (Visser et al., 2021) – and, potentially, reviewed by CPR – is medical research, rather than a greater propensity for medical research to appear in questionable journals. Regardless, the predominance of medical research in questionable journals is concerning given that medical practitioners rely on current publications to inform their treatment decisions and many cannot recognise questionable journals (Richtig et al., 2019).

Overall, MHS articles in questionable journals were significantly less impactful than MHS articles in non-questionable journals. However, research in questionable journals was still extensively cited. The majority of articles had been cited at least once within 3 years of publication in nearly 2,000 non-questionable journals by authors from a diverse array of countries, with only 12% remaining uncited. This was slightly more than the uncited rate of articles in non-questionable journals (8%), but substantially lower than other samples of research in questionable journals that found 25% to 56% were uncited (Bagues et al., 2019; Björk et al., 2020). The uptake of content in questionable journals cannot be attributed to self-citations, which accounted for only 12% of citations and was actually lower than the self-citation rate observed for non-questionable journals (16%), although higher than the 9.5% observed in another sample (Kulczycki et al., 2021). Further, the articles here received on average 4.6 citations each, which is notably more than the mean or median of 0.04-2.6 citations observed in several other studies (Akça & Akbulut, 2021; Bagues et al., 2019; Björk et al., 2020; Ezinwa Nwagwu & Ojemeni, 2015; Frandsen, 2017). The exception is a study of questionable marketing journals in which articles received an average of 6.2 citations (Moussa, 2020). In this case, the questionable journals' mimicking of reputable journals' titles may have influenced the citations of these articles. The higher mean citations and lower uncited rate in this study may also reflect citation practices in the MHS field (Stephen & Stahlschmidt, 2021) or a potential citation advantage afforded to researchers in Germany, as a high-income European country (Akre et al., 2009). However, it demonstrates that questionable research is widely cited in mainstream research channels.

The sharp curtail of international collaboration and the majority of authors in questionable journals being affiliated with hospitals and medical clinics suggests these authors may primarily be medical practitioners. Practitioners working in private practices and non-university hospitals are less acquainted with questionable journals than non-practitioner colleagues engaged in more purely academic pursuits (Richtig et al., 2019) and therefore may be more susceptible to publishing in these journals. Informational campaigns may then reduce the amount of research appearing in questionable journals, as was perhaps observable with the simultaneous decrease in German-authored articles in these venues as publications about predatory publishing rose in 2015-6. However, it is not always the case that research mistakenly appears in questionable journals. Indeed, the participation by some researchers in this dubious publication practice spurred the shift in terminology from "predatory", which portrayed researchers as victims, to "questionable" to recognise researchers' sometimes active role. A competitive, publish or perish culture where quantity of publications is often valued and rewarded over quality is a strong motivating factor for researchers to engage with questionable journals offering fast publishing timelines and no risk of rejection or lengthy revision processes (Frandsen, 2019).

This culture may have implications for the quality of articles appearing in questionable journals. However, the substantial uptake observed here of research in questionable journals by authors in a broad set of countries and journals may suggest that the quality of the research is acceptable, despite its publishing venue. Although the journals have displayed questionable practices and the articles may be inadequately peer-reviewed, this does not necessarily reflect the quality of the article and underlying research itself. For instance, Oermann et al. (2018) rated the quality of over 300 articles in questionable nursing journals based on factors such as their methodological appropriateness, and validity of conclusions and found that, although 48% were of poor quality, 4% could be considered of excellent quality, while the remainder were rated average. Consequently, much broader systematic changes are required to address the research and publishing culture that motivates researchers to use these venues to reduce the publication rate of poor quality research and to shift research of acceptable quality to reputable venues that will not cast doubt on its validity. Currently, further research is necessary to identify what and how such changes could effectively be implemented to address questionable publishing practices.

**Limitations**

This study has several strengths. For example, it has one of the larger samples sizes in this area of research and appears to be among the first to examine the profile, impact, and reach of articles in questionable journals authored by researchers in Germany. It also examined the type of institution of the authors. However, there are also some limitations to consider. First, further examinations have since identified that the metadata for questionable journals is often incomplete in Dimensions, which likely limited the number of articles that were identifiable as authored by researchers in Germany. As such, this study likely underestimates the total proportion of research appearing in questionable journals. Further, I did not examine the content of citations and thus cannot deduce the purpose for citing articles in questionable journals. Consequently, the citations may be to disagree with, refute, or highlight inconsistencies with such articles rather than to support them or integrate their findings. However, in a study of nearly 10,000 citations from WoS journals to articles in questionable journals, negative citations constituted less than 1% of all citations and 97% were neutral (Taskin et al., 2022), so this seems unlikely.